\documentclass[journal ]{new-aiaa}
\usepackage[utf8]{inputenc}
\usepackage{textcomp}
\usepackage{graphicx}
\usepackage{amsmath}
\usepackage[version=4]{mhchem}
\usepackage{longtable,tabularx}
\usepackage{graphicx}
\usepackage{amsmath}
\usepackage{placeins}
\usepackage{subfigure}
\usepackage{algorithmic}
\usepackage{algorithm}
\graphicspath{{./../../Figures/}}
\usepackage{xcolor,colortbl}
\allowdisplaybreaks

\title{Logspace Sequential Quadratic Programming for Design Optimization}

\author{Cody J. Karcher \footnote{Graduate Student, Department of Aeronautics and Astronautics, ckarcher@mit.edu, AIAA Member} }
\affil{Massachusetts Institute of Technology, Cambridge, MA, 02139}

\begin{document}
\maketitle

\begin{abstract}
A novel approach to exploiting the log-convex structure present in many design problems is developed by modifying the classical Sequential Quadratic Programming (SQP) algorithm.  The modified algorithm, Logspace Sequential Quadratic Programming (LSQP), inherits some of the computational efficiency exhibited by log-convex methods such as Geometric Programing and Signomial Programing, but retains the the natural integration of black box analysis methods from SQP.  As a result, significant computational savings is achieved without the need to invasively modify existing black box analysis methods prevalent in practical design problems.  In the cases considered here, the LSQP algorithm shows a 40-70\% reduction in number of iterations compared to SQP.
\end{abstract}

\section{Introduction and Motivation}

A Geometric Program (GP) is a specific type of non-linear optimization problem that becomes convex upon a transformation to logspace\footnote{In some of the literature, the transformation considered in this work is referred to as a log-log transformation since both dependent and independent variables are transformed.  In all cases here, logspace, log-convexity, log transformation etcetera could equivalently be called log-log space, log-log convexity, log-log transformation and similar.} \cite{boyd2007tutorial}.  GPs have received a great deal of attention in many industries including chemical engineering \cite{clasen1984solution}, environment quality control \cite{greenberg1995mathematical}, digital circuit design \cite{boyd2005digital}, analog and RF circuit design \cite{boyd2001optimal,li2004robust,xu2004oracle}, transformer design \cite{jabr2005application}, communication systems \cite{chiang2005geometric,chiang2007power,kandukuri2002optimal}, biotechnology \cite{marin2007optimization,vera2010optimization}, epidemiology \cite{preciado2014optimal}, optimal gas flow \cite{misra2014optimal}, and tree-water-network control \cite{sela2015control} (compilation from Agrawal \cite{agrawal2019disciplined}), but the most recent surge has occurred in the field of aircraft design \cite{hoburg2014geometric,torenbeek2013advanced,hoburg2013fast,kirschen2016signomial,brown2018vehicle,york2018efficient,burton2018solar,lin2020simultaneous,kirschen2018application,york2018turbofan,saab2018robust,hall2018assessment}. 

Geometric Programs are attractive for aircraft design for two reasons.  First, as convex programs GPs can be solved rapidly and with guaranteed convergence to a global optimum.  Second, though all convex programming formulations only allow an objective and constraints of a specific form, the GP formulation happens to be well suited to aircraft design problems \cite{hoburg2014geometric}.  

In addition, Signomial Programming (SP) is a natural extension of the GP formulation that allows for an objective and constraints which are non-convex \cite{boyd2007tutorial}, enabling work by Kirschen \cite{kirschen2016signomial} and York \cite{york2018efficient} to consider the design of subsonic transport aircraft with more than 500 design variables.  But despite the enhanced modeling capability, both the Geometric and Signomial Programming formulations often remain too limiting for designers, particularly those in industry.  Hall \cite{hall2018assessment} specifically notes that both GP and SP formulations require all models to be explicitly written as constraints in the optimization formulation, which effectively eliminates the use of black box analysis models that are prevalent in practical aircraft design \cite{martins2013multidisciplinary}.  

Due to the existence of these black boxes, Sequential Quadratic Programming (SQP) has become one of the more popular algorithms for aircraft design applications \cite{wakayama1998multidisciplinary, kroo1988quasi}.  With SQP, the objective and each constraint are approximated using only a function evaluation and the gradient vector at a series of candidate points.  For black boxes constructed to return gradients (as has become standard practice), the integration with SQP is quite natural.  

This work builds on these two fundamental pillars.  First, efforts in Geometric and Signomial Programming have revealed an exploitable convex underlying structure in engineering design, but these formulations do not conform to the existing modeling approach.  Furthermore, modifying existing models would require an investment of time and money that would likely be prohibitive.  Second, the SQP algorithm conforms with existing modeling approaches but ignores known structure in the optimization formulation, causing valuable computational effort to be wasted in the form of unnecessary iterations.  The modification to SQP proposed in this work, Logspace Sequential Quadratic Programming (LSQP), bridges this divide to exploit underlying design space structure while remaining practical to designers who wish to continue utilizing their existing methods and practices.
\section{Foundations in Established Optimization Methods}
\subsection{Sequential Quadratic Programming}
Consider the general non-linear constrained optimization problem:
\begin{equation}
    \begin{aligned}
        & \underset{\mathbf{x}}{\text{minimize}}
        & & f(\mathbf{x})\\
        & \text{subject to}
        & & \mathbf{g}_i (\mathbf{x}) \leq 0 , \; i = 1, \ldots, N\\
        &&& \mathbf{h}_j (\mathbf{x}) = 0 , \; j = 1, \ldots, M
    \end{aligned}
    \label{NLP}
\end{equation}
The Sequential Quadratic Programming (SQP) algorithm solves this general non-linear program (NLP) by iteratively considering candidate solutions $\mathbf{x}_k$ and solving the Quadratic Programming (QP) sub-problem:
\begin{equation}
    \begin{aligned}
        & \underset{\mathbf{d}}{\text{minimize}}
        & & f(\mathbf{x}_k) + 
            \nabla f(\mathbf{x}_k)^T \mathbf{d} + 
            \frac{1}{2} \mathbf{d}^T    \nabla^2 \mathcal{L} (\mathbf{x}_k)   \mathbf{d}\\
        & \text{subject to}
        & & g_i(\mathbf{x}_k) + \nabla g_i(\mathbf{x}_k)^T \mathbf{d} \leq 0 , \; i = 1, \ldots, N\\
        &&& h_j(\mathbf{x}_k) + \nabla h_j(\mathbf{x}_k)^T \mathbf{d}    = 0 , \; j = 1, \ldots, M \\
        &&& \mathbf{d} = \mathbf{x}-\mathbf{x}_k 
    \end{aligned}
    \label{SQP_subproblem}
\end{equation}
at each iteration $k$ \cite{boggs1996sequential,nocedal2006numerical,kraft1988software} until some convergence criteria is reached.  The sub-problem is constructed by approximating the objective as a second order Taylor Series about the candidate solution $\mathbf{x}_k$\footnote{Note that the use of the Hessian of the Lagrangian $\nabla^2 \mathcal{L}$ has been shown to be superior to directly using the Hessian of the objective function $\nabla^2 f$ \cite{boggs1996sequential}.} and each constraint as a first order Taylor Series about the same $\mathbf{x}_k$.  This process results in a convex QP that can be readily solved for a global optimum $\mathbf{d}^*$, the direction vector that points from current point $\mathbf{x}_k$ to the next point $\mathbf{x}_{k+1}$.  The optimal step size is then computed via a line search procedure.  

Note that this sub-problem can be fully constructed given only the function evaluations $f(\mathbf{x}_k)$, $g_i(\mathbf{x}_k)$, and $h_j(\mathbf{x}_k)$ and the gradients $\nabla f(\mathbf{x}_k)$, $\nabla g_i(\mathbf{x}_k)$, and $\nabla h_j(\mathbf{x}_k)$, which is the key to integrating existing black box analysis models.
\subsection{Geometric Programming}
Geometric Programs are built upon two fundamental building blocks: monomial and posynomial functions.  A monomial function is defined as the product of a leading constant with each variable raised to a real power \cite{boyd2007tutorial}:
\begin{equation}
    m(\textbf{x}) = c{x_1}^{a_1}{x_2}^{a_2}...\;{x_n}^{a_n} = c \prod_{i=1}^{N} x_i^{a_i}
\end{equation}
A posynomial is simply the sum of monomials \cite{boyd2007tutorial}, which can be defined in notation as:
\begin{equation}
  p(\textbf{x}) = m_1(\textbf{x}) +  m_2(\textbf{x}) + ... + m_n(\textbf{x}) = \sum_{k=1}^{K} c_k \prod_{i=1}^{N} x_i^{a_{ik}}
\end{equation}
From these two building blocks, it is possible to construct the definition of a GP in standard form \cite{boyd2007tutorial}:
\begin{equation}
    \begin{aligned}
        & \underset{\mathbf{x}}{\text{minimize}}
        & & p_0 (\textbf{x}) \\
        & \text{subject to}
        & & m_i (\textbf{x}) = 1, \; i = 1, \ldots, N\\
        &&& p_j (\textbf{x}) \leq 1, \; j = 1, \ldots, M
    \end{aligned}
    \label{GP_standard}
\end{equation}
When constraints and objectives can be written in the form specified in Equation \ref{GP_standard} it is said that the problem is \textit{GP compatible}.

In general, the formulation defined by Equation \ref{GP_standard} is a non-linear and non-convex optimization problem, making it extremely difficult to solve \cite{boyd2007tutorial}, however a GP can be transformed into a convex optimization problem by undergoing a logarithmic transformation.  As convex programs GPs can be solved by a wide variety of algorithms, but most are now solved using primal/dual methods \cite{boyd2007tutorial} and solvers such as CVXOPT \cite{andersen2013cvxopt} are readily available and return reliable results under most circumstances.

\subsection{Signomial Programming} \label{sig_sec}
Signomal Programs (SPs) are a logical extension of Geometric Programs that enable the inclusion of negative leading constants and a broader set of equality constraints.  Of interest here is that SPs are \textit{not} convex upon transformation to logspace unlike their GP counterparts, but still benefit from an underlying structure which is well approximated by a log-convex formulation.  This property is what is referred to here has having a high degree of log-convexity.  

The key building blocks of the Signomial Programing are signomials, which are the difference between two posynomials $p(\textbf{x})$ and $n(\textbf{x})$:
\begin{equation}
  s(\textbf{x}) = p(\textbf{x}) -  n(\textbf{x}) = \sum_{k=1}^{K} c_k \prod_{i=1}^{N} x_i^{a_{ik}} - \sum_{p=1}^{P} d_p \prod_{i=1}^{N} x_i^{g_{ik}}
\end{equation}
where posynomial $p(\textbf{x})$ represents the convex portion of the signomial and `neginomial' $n(\textbf{x})$ is the concave portion.  With this definition, it is now possible to write the standard form for a Signomial Program \cite{kirschen2016signomial}:
\begin{equation}
    \begin{aligned}
        & \underset{\mathbf{x}}{\text{minimize}}
        & & \frac{p_0 (\textbf{x})}{n_0 (\textbf{x})} \\
        & \text{subject to}
        & & s_i (\textbf{x}) = 0, \; i = 1, \ldots, N\\
        &&& s_j (\textbf{x}) \leq 0, \; j = 1, \ldots, M
    \end{aligned}
    \label{SP_standard}
\end{equation}
however, another useful form is:
\begin{equation}
    \begin{aligned}
        & \underset{\mathbf{x}}{\text{minimize}}
        & & \frac{p_0 (\textbf{x})}{n_0 (\textbf{x})} \\
        & \text{subject to}
        & & \frac{p_i (\textbf{x})}{n_i (\textbf{x})} = 1, \; i = 1, \ldots, N\\
        &&& \frac{p_j (\textbf{x})}{n_j (\textbf{x})} \leq 1, \; j = 1, \ldots, M
    \end{aligned}
    \label{SP_standardMod}
\end{equation}
In this alternative form, the neginomial is added to both sides, and then used as a divisor to construct an expression either equal to or constrained by a value of one. 

The Difference of Convex Algorithm (DCA) is used to solve SPs.  In this method, the neginomials are replaced by their monomial approximation \cite{boyd2007tutorial}:
\begin{equation}
  \begin{aligned}
      \bar{n}(\mathbf{x})|_{\mathbf{x}_k} & = n(\mathbf{x}_k) \prod_{i=1}^{N} \left( \frac{x_i}{x_{k_i}}  \right)^{a_i} \\
      a_i & = \frac{x_{k_i}}{n(\mathbf{x}_k)} \frac{\partial n}{\partial x_i}
  \end{aligned}
  \label{monomialApprox}
\end{equation}
thus yielding a GP at each iteration:
\begin{equation}
    \begin{aligned}
        & \underset{\mathbf{x}}{\text{minimize}}
        & & \frac{p_0 (\textbf{x})}{\bar{n}_0 (\textbf{x})} \\
        & \text{subject to}
        & & \frac{p_i (\textbf{x})}{\bar{n}_i (\textbf{x})} = 1, \; i = 1, \ldots, N\\
        &&& \frac{p_j (\textbf{x})}{\bar{n}_j (\textbf{x})} \leq 1, \; j = 1, \ldots, M
    \end{aligned}
    \label{SP_gpsub}
\end{equation}

It has been shown in many test cases \cite{kirschen2016signomial,brown2018vehicle,york2018efficient,burton2018solar,lin2020simultaneous,kirschen2018application,york2018turbofan} that extremely complex Signomial Programs can be constructed which exhibit a high degree of log-convexity and can therefore be solved in very few iterations.  For example, Kirschen \cite{kirschen2016signomial} proposes a full aircraft design problem with 824 variables that solves in only six iterations.  Similarly, the design case presented by York \cite{york2018turbofan} has 628 variables and also solves in only six iterations.
\section{Mathematical Definition of Logspace Sequential Quadratic Programming} \label{mathdef}
Consider now a slight adjustment to Equation~\ref{NLP}:
\begin{equation}
    \begin{aligned}
        & \underset{\mathbf{x}}{\text{minimize}}
        & & f(\mathbf{x})\\
        & \text{subject to}
        & & \mathbf{g}_i (\mathbf{x}) \leq 1 , \; i = 1, \ldots, N\\
        &&& \mathbf{h}_j (\mathbf{x}) = 1 , \; j = 1, \ldots, M
    \end{aligned}
    \label{lsqpsf}
\end{equation}
reformulated so that a value of 1 appears on the right hand side of the constraints.  

From Equation~\ref{lsqpsf}, the problem can now be transformed in a similar fashion to that of a Geometric Program \cite{boyd2007tutorial}.  Taking the transformation $y_i = \log{x_i}$, or equivalently $x_i = e^{y_i}$, the transformed problem becomes:
\begin{equation}
    \begin{aligned}
        & \underset{\mathbf{y}}{\text{minimize}}
        & & \log{ f (e^{\textbf{y}}) } \\
        & \text{subject to}
        & & \log{ \mathbf{g}_i (e^{\textbf{y}}) } \leq 0, \; i = 1, \ldots, N\\
        &&& \log{ \mathbf{h}_j (e^{\textbf{y}}) } = 0, \; j = 1, \ldots, M
    \end{aligned}
    \label{NLP_logtransformed}
\end{equation}
Consider that rather than implementing the SQP algorithm on the original problem (Equation~\ref{NLP}), that the same SQP algorithm can be implemented on this transformed problem (Equation~\ref{NLP_logtransformed}).  Derivatives will be necessary for this modified SQP, and are provided by Boyd \cite{boyd2007tutorial}:
\begin{equation}
    \frac{\partial \log{ f (e^{\textbf{y}}) }}{\partial y_i} = \frac{x_i}{f(\mathbf{x})} \frac{\partial f}{\partial x_i} 
    \label{gradient_F}
\end{equation}
and are similar for the functions $g_i(e^{\textbf{y}})$ and $h_j(e^{\textbf{y}})$.  The SQP sub-problem then becomes:
\begin{equation}
    \begin{aligned}
        & \underset{\mathbf{d}}{\text{minimize}}
        & & \log{f(\mathbf{x}_k)} + 
            \frac{1}{f(\mathbf{x}_k)}\left( \mathbf{x}_k \odot \nabla f(\mathbf{x}_k) \right)^T \mathbf{d} + 
            \frac{1}{2} \mathbf{d}^T  \nabla^2 \mathcal{L}(\mathbf{y}_k)   \mathbf{d}\\
        & \text{subject to}
        & & \log{g_i(\mathbf{x}_k)} + \frac{1}{g_i(\mathbf{x}_k)}\left( \mathbf{x}_k \odot \nabla g_i(\mathbf{x}_k) \right)^T \mathbf{d} \leq 0, \; i = 1, \ldots, N\\
        &&& \log{h_j(\mathbf{x}_k)} + \frac{1}{h_j(\mathbf{x}_k)}\left( \mathbf{x}_k \odot \nabla h_j(\mathbf{x}_k) \right)^T \mathbf{d}    = 0 , \; j = 1, \ldots, M \\
        &&& \mathbf{d} = \mathbf{y}-\log{\mathbf{x}_k} \\
        &&& \mathbf{y} = \log{\mathbf{x}}
    \end{aligned}
    \label{lsqp_sp}
\end{equation}
The proposed method, Logspace Sequential Quadratic Programming (LSQP), solves the general non-linear program (Equation~\ref{lsqpsf}) by iteratively solving a series of approximate sub-problems (Equation~\ref{lsqp_sp}) that are quadratic programs under a log transformation.

If foreknowledge of log convexity is assumed, Equation~\ref{NLP_logtransformed} is expected to be well approximated as a log convex optimization problem based on the similar properties of geometric and signomial programs.  For example, in the case where the original NLP is a geometric program, monomial constraints will be exactly represented as lines in logspace, while posynomials be represented as log-sum-exp functions which are known to be convex for positive leading constants.  Thus, the quadratic programming approximation after log transformation (Equation~\ref{lsqp_sp}) should be a more accurate representation of the original NLP than the quadratic programming approximation with no transformation (Equation~\ref{SQP_subproblem}).  This superior representation should yield computational savings, and indeed it does.
\section{The LSQP Algorithm} \label{lsqp_algo}
LSQP can be implemented in two ways.  The simplest approach is to write the problem in the modified standard form (Equation~\ref{lsqpsf}), apply the log transformation (Equation~\ref{NLP_logtransformed}), and then solve iteratively by using a series of standard SQP sub-problems (Equation~\ref{SQP_subproblem}).  The final solution $\textbf{x}^*$ is then obtained by a simple reverse transformation of $\textbf{y}^*$.  The benefit of this approach is that existing SQP algorithms can be used to solve the problem, an approach demonstrated by Kirschen \cite{kirschen2018power} in limited form.  But this method only works well if all of the functions $f(\textbf{x})$, $g_i(\textbf{x})$, and $h_j(\textbf{x})$ are known explicitly, and if the optimization framework enables the log transformations to be easily implemented.  Additionally, the practical numeric considerations of log-sum-exp functions \cite{hoburg2016data} may cause issues in this type of implementation.

A more tailored algorithm that is based on an understanding of the underlying mathematics of LSQP has four key benefits.  First is usability as a practical algorithm.  For software tools like Matlab and Python, it is important for a user to be able to swap between algorithms with minimal effort.  A tailored algorithm hides the log transformation `under the hood' which facilitates easy integration with existing frameworks.  

Second, true black boxes can be more easily integrated.  Constructing SQP sub-problems from Equation~\ref{NLP_logtransformed} will require gradients $\partial \log{ f (e^{\textbf{y}}) } / \partial y_i$, which have been eliminated in Equation~\ref{lsqp_sp} using Equation~\ref{gradient_F}.  Black boxes can therefore be directly integrated into the log quadratic sub-problem of Equation~\ref{lsqp_sp} without modification, which constitutes a significant advantage.

Third, the construction of standard form for LSQP is not always straightforward.  Consider even the simple constraint $y=x$.  A natural construction of standard form might be to first zero out the constraint and then add one, resulting in $y-x+1 = 1$.  But $x/y=1$ is a far superior constraint since it is linear after the log transformation.  Allowing the algorithm to test for such structure enables higher quality solutions to be produced more frequently.

Fourth, modifications to the standard SQP algorithm may be necessary in future work to improve the algorithm.  Consider that while iteration points $\mathbf{x}_k$ need not be feasible, the conditions $f(\mathbf{x}_k) > 0$, $g_i(\mathbf{x}_k) > 0$, and $h_j(\mathbf{x}_k) > 0$ must all hold true at the initial condition and at each subsequent $\mathbf{x}_k$.  As will be discussed in the section on results, this is most easily enforced by modifying the line search phase of the traditional SQP algorithm to enforce these conditions.  Such modification is not generally realistic for off the shelf SQP algorithms without significant expertise and effort.

Despite the differences, the close relationship between SQP and LSQP means that the existing SQP literature can be heavily utilized.  The algorithm proposed here (Algorithm \ref{alg:lsqp}) is a combination of methods proposed by Nocedal and Wright \cite{nocedal2006numerical}, Kraft \cite{kraft1988software}, and the Matlab documentation \cite{matlab2020constrained}.  These sources seem to form the core of the some of the most commonly used algorithms, such as those implemented in Matlab and Python's Scipy package.  

\begin{algorithm}
\caption{Logspace Sequential Quadratic Programming (LSQP)}
\label{alg:lsqp}
\begin{algorithmic}[1]
\STATE{Given $\mathbf{x}_0$}
\STATE{Construct standard form}
\STATE{Compute $\mathbf{y}_0 = \log ( \mathbf{x}_0 )$}
\STATE{Initialize logspace Lagrange multipliers, $\mu_0 \leftarrow \mathbf{1}$}
\STATE{Initialize the matrix $\mathbf{B} \leftarrow \mathbf{I}$ the approximation of $\nabla^2 \mathcal{L}(\mathbf{y},\mu)$ \cite{nocedal2006numerical}}
\STATE{Compute $f(x_0)$, $g_i(x_0)$, and $h_j(x_0)$ }
\STATE{Compute $\log f(x_0)$, $\log g_i(x_0)$, and $\log h_j(x_0)$ }
\STATE{Compute $\nabla f(x_0)$, $\nabla g_i(x_0)$, and $\nabla h_j(x_0)$}
\STATE{Compute $\nabla \log f(e^{y_0})$, $\nabla \log g_i(e^{y_0})$, and $\nabla \log h_j(e^{y_0})$ via Equation \ref{gradient_F}}
\FOR{$k=0$ to maxIter}
  \STATE{Solve the QP sub-problem to obtain $d_y$ and $d_{\mu}$ \cite{nocedal2006numerical}}
  \STATE{Compute the step size $\alpha_k$ via inexact line search \cite{nocedal2006numerical,matlab2020constrained}}
  \STATE{$\mathbf{y}_{k+1} \leftarrow \mathbf{y}_k + \alpha_k d_y$}
  \STATE{$\mathbf{x}_{k+1} \leftarrow \exp(\mathbf{y}_{k+1})$}
  \STATE{$\mu_{k+1} \leftarrow \mu_k + \alpha_k d_{\mu}$}
  \STATE{Compute $f(x_{k+1})$, $g_i(x_{k+1})$, and $h_j(x_{k+1})$}
  \STATE{Compute $\log f(x_{k+1})$, $\log g_i(x_{k+1})$, and $\log h_j(x_{k+1})$}
  \STATE{Compute $\nabla f(x_{k+1})$, $\nabla g_i(x_{k+1})$, and $\nabla h_j(x_{k+1})$}
  \STATE{Compute $\nabla \log f(e^{y_{k+1}})$, $\nabla \log g_i(e^{y_{k+1}})$, and $\nabla \log h_j(e^{y_{k+1}})$ via Equation \ref{gradient_F}}
  \STATE{Compute $\nabla \mathcal{L}_k(\mathbf{y}_k,\mu_{k+1})$ \cite{nocedal2006numerical}}
  \STATE{Compute $\nabla \mathcal{L}_{k+1}(\mathbf{y}_{k+1},\mu_{k+1})$ \cite{nocedal2006numerical}}
  \IF{$\nabla \mathcal{L}_{k+1}(\mathbf{y}_{k+1},\mu_{k+1}) < \varepsilon_{GL}$ \cite{nocedal2006numerical}}
    \RETURN{$x_{k+1}$}
  \ELSIF{$|| d_{x} || < \varepsilon_{dx}$ \cite{matlab2020constrained}}
    \RETURN{$x_{k+1}$}
  \ELSE
    \STATE{Perform a damped BFGS update on matrix $\mathbf{B}$ \cite{nocedal2006numerical}}
    \STATE{$k \leftarrow k+1$}
  \ENDIF
\ENDFOR
\RETURN{$x_k$, maximum iteration count reached}
\end{algorithmic}
\end{algorithm}
\section{Comparison of the Performance of the SQP and LSQP Algorithms}
\subsection{Methodology} 
Four test problems were taken from the literature to test the performance of LSQP against the classical SQP algorithm, named after the author who originally proposed the problem: Boyd \cite{boyd2007tutorial}, Rosenbrock, Floudas \cite{floudas2013handbook}, and Kirschen-Ozturk \cite{kirschen2018power}.  For each test problem, four algorithms were run from a set of common randomly sampled initial guesses:
\begin{itemize} 
    \item A Python implementation of SQP (implemented to be identical to the LSQP algorithm without the log transformation) 
    \item A Matlab implementation of SQP (fmincon with the `SQP' flag)  
    \item A Python implementation of LSQP (as described in Section \ref{lsqp_algo})
    \item A log transformation followed by application of Matlab SQP (abbreviated as LT+SQP)
\end{itemize}
The SQP algorithm in Python's Scipy package was also considered for comparison (an implementation of the algorithm from Kraft \cite{kraft1988software}), but lack of control on the termination condition meant a meaningful comparison was not possible.  


For each of the 16 problem/algorithm combinations, 2000 trials were run starting from a random initial starting point.  In 1000 of these cases, the initial guess was bounded to be within +/- 10\% of the known optimum, and for the other 1000 the initial guess was bounded to be within +/-80\% of the known optimum.  These are referred to as a `Good' and `Poor' initial guess respectively in the following sections.

Due to the computational expense of 32000 trials, the computational resources of the MIT SuperCloud were utilized \cite{reuther2018interactive}, and a limit of 500 iterations was placed on all 4 algorithms.

In all four test problems the objective and constraints were known explicitly, and so analytical derivatives were used in the two Python cases.  Both cases using the Matlab SQP algorithm were implemented with the default gradient computation method, as these cases were not the focus of this work.  Note that none of these cases contains a true black boxed analysis model in order to allow for comparison between known optima and the computed result, however the objective and constraints are treated by the algorithm (Algorithm \ref{alg:lsqp}) as black boxes regardless of actual form and therefore any equivalent black box could be substituted in without affecting the optimizer in any way, so long as gradients are somehow made available.  
\subsection{Results from the Boyd Geometric Program}
The first test case is a toy problem proposed by Boyd as an example of a simple Geometric Program \cite{boyd2007tutorial}, given here in LSQP standard form:
\begin{equation}
\begin{aligned}
     \underset{h,w,d}{\text{minimize}} \quad &  1/(hwd) \\
     \text{subject to}  \quad & 2 \frac{hw}{A_{wall}} + 2 \frac{hd}{A_{wall}} \leq 1 \\
    & \frac{wd}{A_{floor}} \leq 1 \\
    & \frac{\alpha w}{h} \leq 1 \\
    & \frac{h}{\beta w} \leq 1 \\
    & \frac{\gamma w}{d} \leq 1 \\
    & \frac{d}{\delta w} \leq 1
\end{aligned}
\label{boyd_gp}
\end{equation}

\begin{figure}[htb]
\centering     
\subfigure[A Good Initial Guess]{\includegraphics[width=0.49\textwidth]{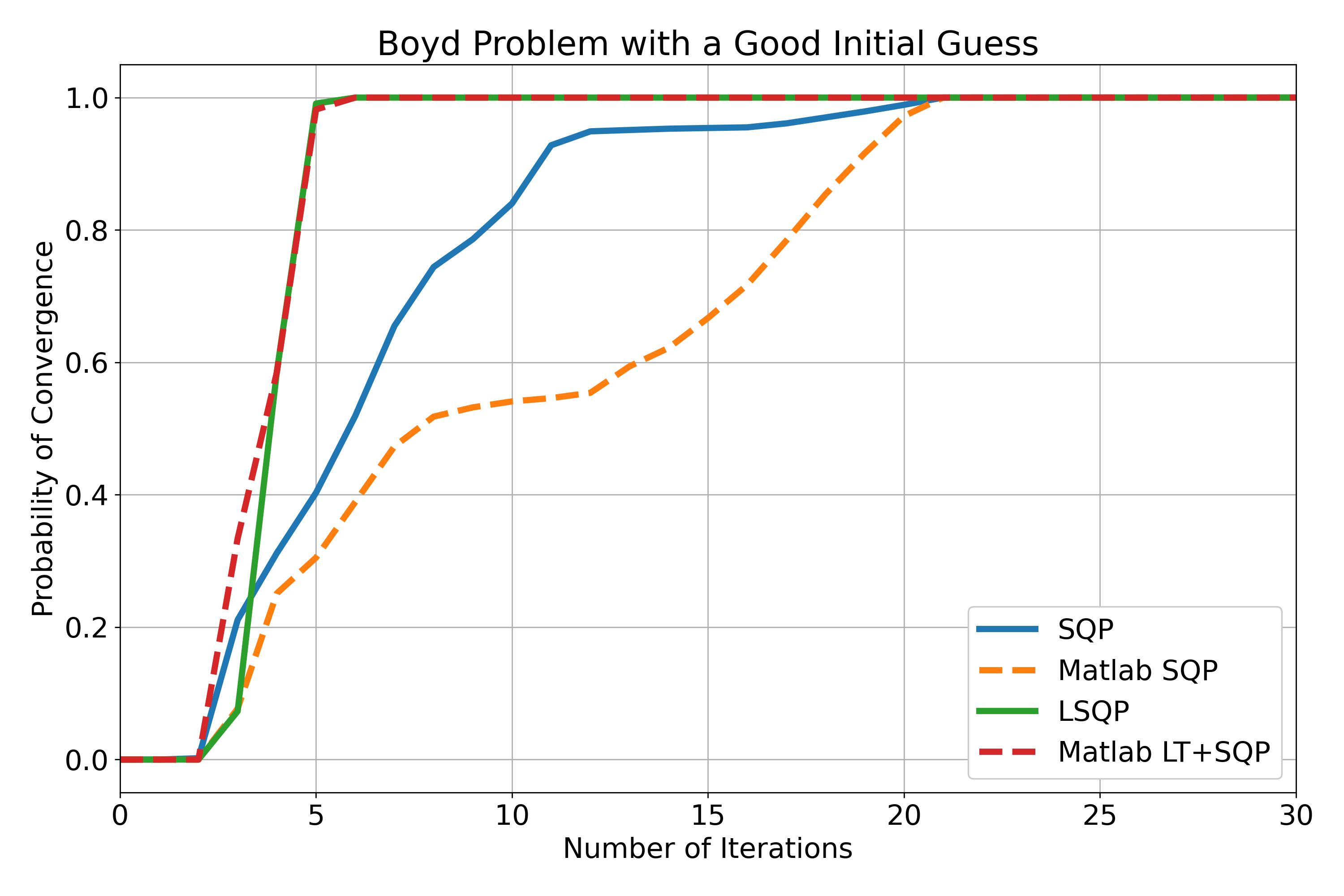}}
\subfigure[A Poor Initial Guess]{\includegraphics[width=0.49\textwidth]{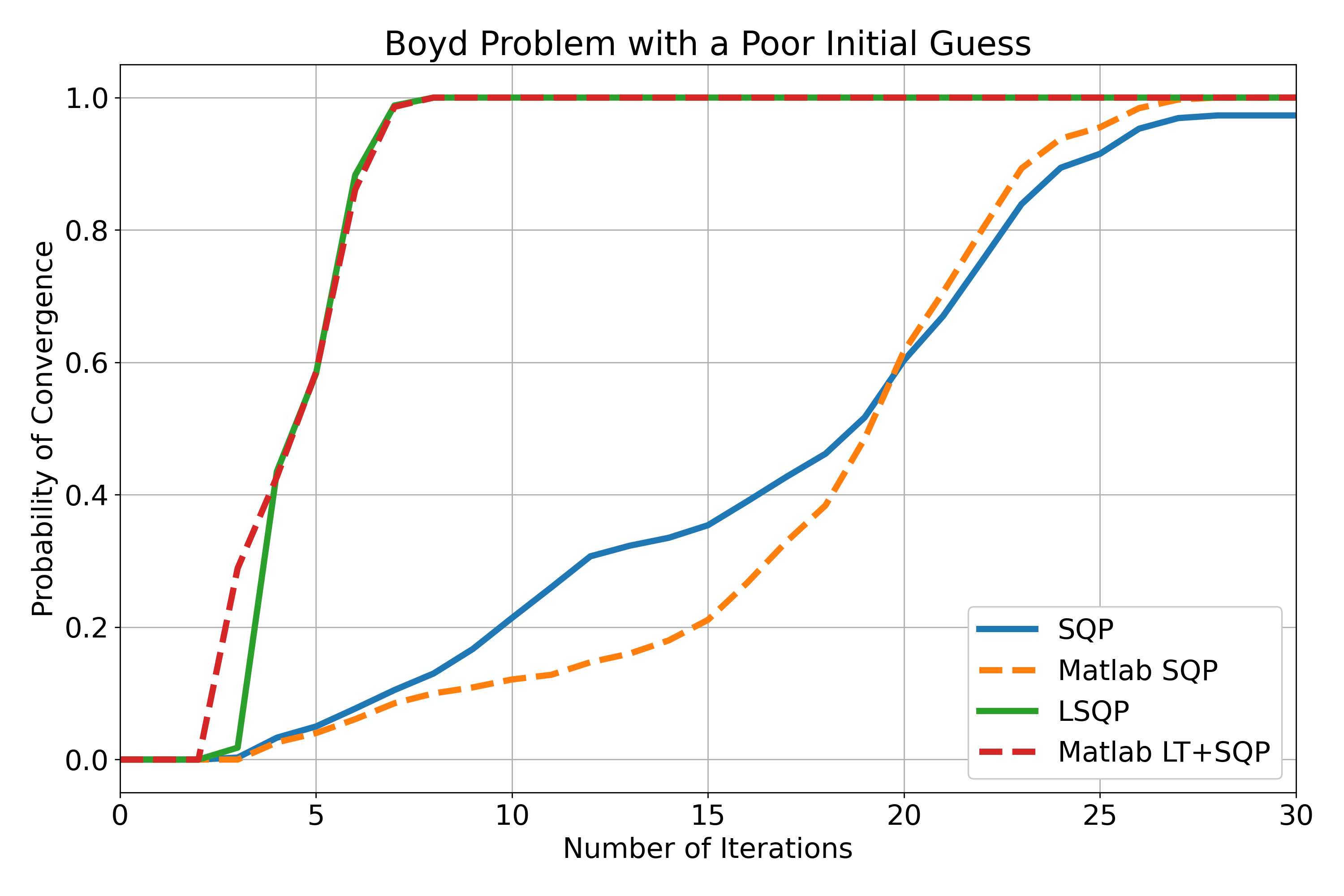}}
\caption{Probability of Convergence vs. Iteration Count for the Boyd Problem}
\label{boyd_results_fig}
\end{figure}

Solving this problem with the LSQP modification requires significantly fewer iterations than traditional SQP.  Figure \ref{boyd_results_fig} shows that all of the LSQP trials converged within 10 iterations, while the best SQP case converges in the same number of iterations less than 80\% of the time for a good initial guess, and less than 20\% of the time for a poor initial guess.  Tables \ref{t:scheme_comparison_Boyd_Good} and \ref{t:scheme_comparison_Boyd_Poor} provide a more detailed breakdown of the data runs.

\begin{table}[htbp] 
  \footnotesize 
  \begin{center} 
  \caption{Results of Solving the Boyd Problem With a Good Initial Guess} 
  \label{t:scheme_comparison_Boyd_Good} 
  \begin{tabular}{ c  c  c  c  c  c } 
      \cline{2-6}
                    & Optimum   & SQP                 & Matlab SQP          & LSQP                & Matlab LT+SQP        \\ 
      \hline 
      Obj $[1/m^3]$ & 5.196e-03 & 5.199e-03 (+0.06\%) & 5.196e-03 (+0.00\%) & 5.196e-03 (-0.00\%) & 5.196e-03 (-0.00\%)  \\ \hline 
      $d$ $[m]$     & 11.55     & 11.27 (-0.17\% )    & 11.53 (-0.17\% )    & 11.55 (+0.00\%)     & 11.55 (+0.00\%)     \\ \hline 
      $h$ $[m]$     & 2.89      & 2.92 (+0.09\%)      & 2.89 (+0.09\%)      & 2.89 (-0.00\%)      & 2.89 (-0.00\%)      \\ \hline 
      $w$ $[m]$     & 5.77      & 5.85 (+0.09\%)      & 5.78 (+0.09\%)      & 5.77 (-0.00\%)      & 5.77 (-0.00\%)      \\ \hline 
      Iterations    & -         & 6.94 (0.00\%)       & 10.68 (+53.99\%)    & 4.35 (-37.31\%)     & 4.10 (-40.90\%)      \\ \hline 
      Failures      & -         & 0 (0.00\%)          & 0 (0.00\%)          & 0 (0.00\%)          & 0 (0.00\%)           \\ \hline 
  \end{tabular} 
 \end{center} 
\end{table} 

\begin{table}[htbp] 
  \footnotesize 
  \begin{center} 
  \caption{Results of Solving the Boyd Problem With a Poor Initial Guess} 
  \label{t:scheme_comparison_Boyd_Poor} 
  \begin{tabular}{ c  c  c  c  c  c } 
      \cline{2-6}
                    & Optimum   & SQP                 & Matlab SQP          & LSQP                & Matlab LT+SQP        \\ 
      \hline 
      Obj $[1/m^3]$ & 5.196e-03 & 5.203e-03 (+0.12\%) & 5.196e-03 (+0.00\%) & 5.196e-03 (-0.00\%) & 5.196e-03 (-0.00\%)  \\ \hline 
      $d$ $[m]$     & 11.55     & 11.36 (-0.05\%)     & 11.54 (-0.05\%)     & 11.55 (+0.00\%)     & 11.55 (+0.00\%)     \\ \hline 
      $h$ $[m]$     & 2.89      & 2.91 (+0.03\%)      & 2.89 (+0.03\%)      & 2.89 (-0.00\%)      & 2.89 (-0.00\%)      \\ \hline 
      $w$ $[m]$     & 5.77      & 5.82 (+0.03\%)      & 5.78 (+0.03\%)      & 5.77 (-0.00\%)      & 5.77 (-0.00\%)      \\ \hline 
      Iterations    & -         & 16.94 (0.00\%)      & 18.27 (+7.84\%)     & 5.09 (-69.94\%)     & 4.85 (-71.37\%)      \\ \hline 
      Failures      & -         & 27 (2.70\%)         & 0 (0.00\%)          & 0 (0.00\%)          & 0 (0.00\%)           \\ \hline 
  \end{tabular} 
 \end{center} 
\end{table} 
\FloatBarrier

The Boyd Geometric Program demonstrates a clear win for LSQP: the number of required iterations is decreased by approximately 70\% at no additional computational cost or sacrifice to solution quality.  This benefit is unsurprising due to the large number of monomials which become exactly represented as affine under the transformation to logspace, and the perfect log-convexity exhibited by all geometric programs.  However, subsequent cases will become more complex and represent greater challenges to the LSQP algorithm.
\subsection{Results from the Constrained Rosenbrock Problem}
The Rosenbrock function is a common test case for optimization methods. It is unique in that the optimum resides in an extremely shallow local valley, making it an excellent test problem for gradient based methods.  Consider the following constrained version of the problem:
\begin{equation}
    \begin{aligned}
         \underset{x,y}{\text{minimize}}  \quad &  (1-x)^2 + 100(y-x^2)^2 + 1\\
         \text{subject to}  \quad & (x-1)^3 - y + 2 \leq 1 \\
        & x + y - 1 \leq 1 \\
        & \frac{x}{1.5} \leq 1 \\
        & \frac{y}{2.5} \leq 1 \\
    \end{aligned}
    \label{rosenbrock}
\end{equation}
Note that a constant of 1 has been added to the objective to shift up the optimum and enable the log transformation of LSQP. The variables are also bound to be greater than a small positive constant to aid the construction of sub-problems.

Rosenbrock's problem is not a GP or SP as formulated, but can be reformulated by setting an intermediate variable $z$ equal to the objective.  However, an attempt to solve the problem in this modified form with the Difference of Convex Algorithm does not succeed.  Fortunately, the optimum is known to be $f(1,1)=1$.  

\begin{figure}[htb]
\centering     
\subfigure[A Good Initial Guess]{\includegraphics[width=0.49\textwidth]{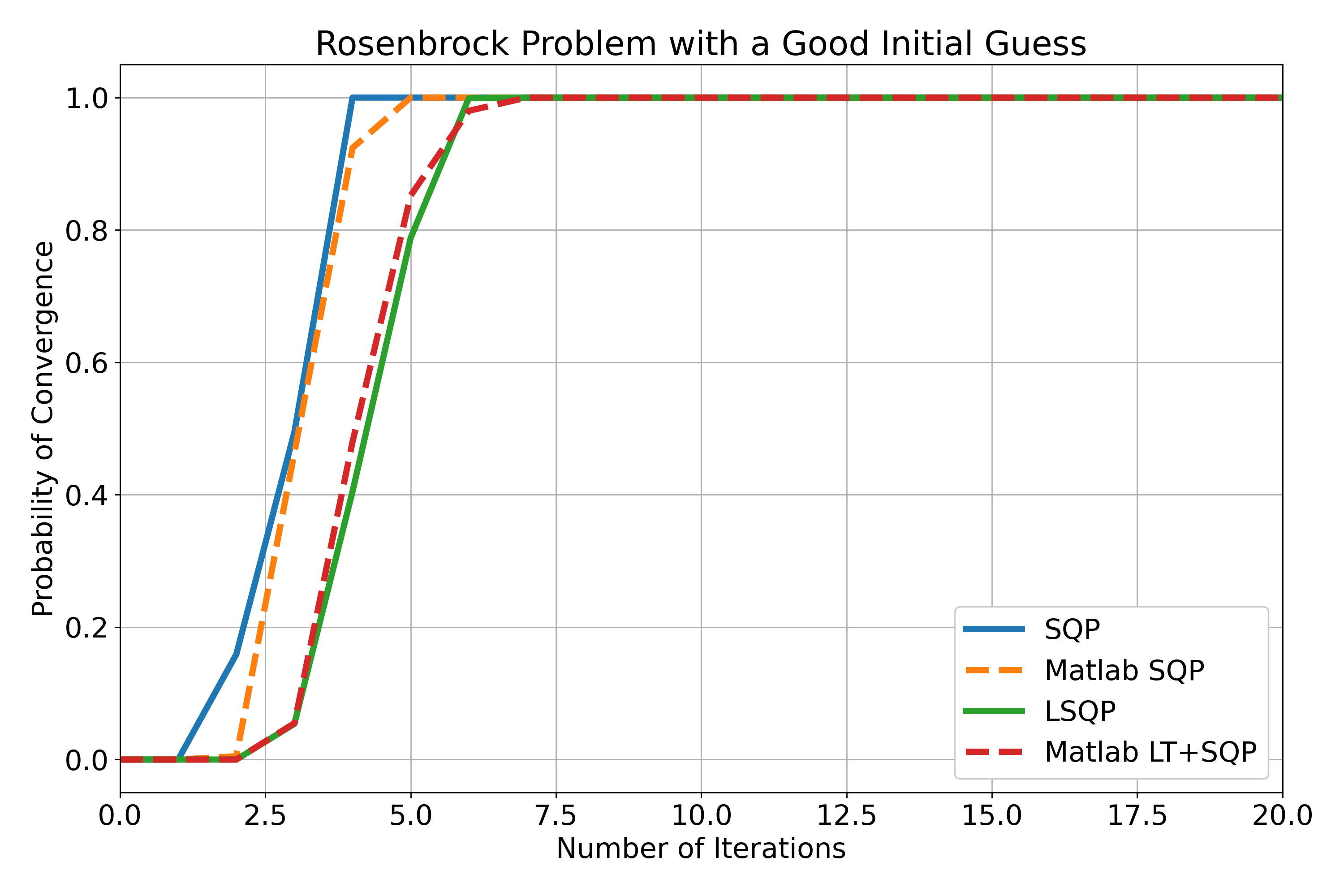}}
\subfigure[A Poor Initial Guess]{\includegraphics[width=0.49\textwidth]{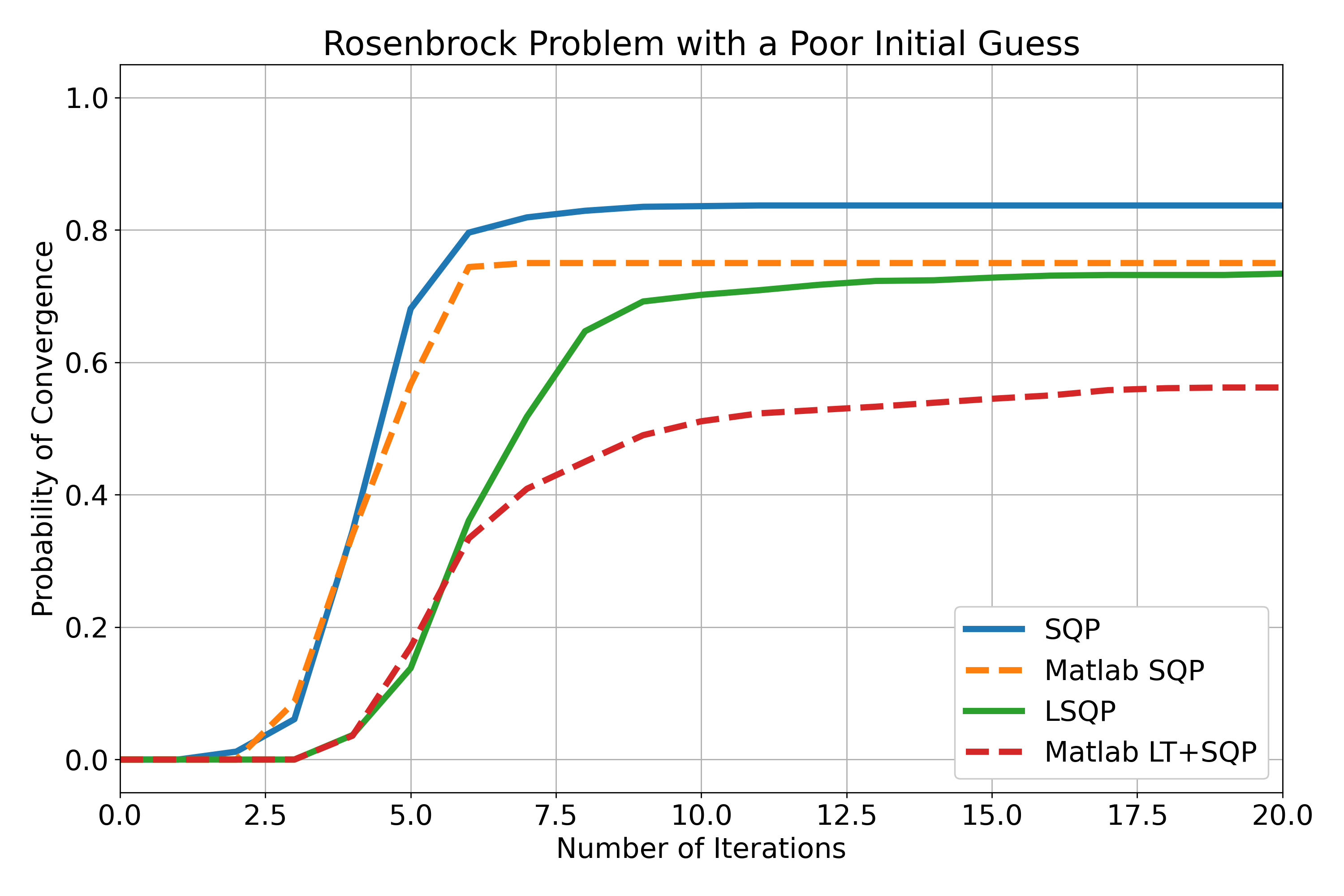} \label{ros_b}}
\caption{Probability of Convergence vs. Iteration Count for the Rosenbrock Problem}
\label{rosenbrock_results_fig}
\end{figure}

Unlike Boyd's Geometric Program, the Rosenbrock problem is a clear win for traditional SQP.  Figure \ref{rosenbrock_results_fig} along with Tables \ref{t:scheme_comparison_Rosenbrock_Good} and \ref{t:scheme_comparison_Rosenbrock_Poor} show a 40-50\% increase in iteration count for the LSQP modification.  

\begin{table}[htbp] 
  \footnotesize 
  \begin{center} 
  \caption{Results of Solving the Rosenbrock Problem With a Good Initial Guess} 
  \label{t:scheme_comparison_Rosenbrock_Good} 
  \begin{tabular}{ c  c  c  c  c  c } 
      \cline{2-6}
                           & Optimum        & SQP & Matlab SQP & LSQP & Matlab LT+SQP    \\ 
      \hline 
      Obj $[-]$ & 1.00 & 1.00 (+0.00\%) & 1.00 (+0.00\%) & 1.00 (+0.00\%) & 1.00 (+0.00\%)  \\ \hline 
      $x$ $[-]$ & 1.00 & 1.00 (-0.00\%) & 1.00 (-0.00\%) & 1.00 (-0.00\%) & 1.00 (-0.00\%)  \\ \hline 
      $y$ $[-]$ & 1.00 & 1.00 (-0.00\%) & 1.00 (-0.00\%) & 1.00 (+0.00\%) & 1.00 (+0.00\%)  \\ \hline 
      Iterations   & - & 3.35 (0.00\%) & 3.61 (+7.84\%) & 4.75 (+42.07\%) & 4.63 (+38.48\%)  \\ \hline 
      Failures & - & 0 (0.00\%) & 0 (0.00\%) & 0 (0.00\%) & 0 (0.00\%)  \\ \hline 
  \end{tabular} 
 \end{center} 
\end{table} 

\begin{table}[htbp] 
  \footnotesize 
  \begin{center} 
  \caption{Results of Solving the Rosenbrock Problem With a Poor Initial Guess} 
  \label{t:scheme_comparison_Rosenbrock_Poor} 
  \begin{tabular}{ c  c  c  c  c  c } 
      \cline{2-6}
                           & Optimum        & SQP & Matlab SQP & LSQP & Matlab LT+SQP    \\ 
      \hline 
      Obj $[-]$ & 1.00 & 1.00 (+0.00\%) & 1.00 (+0.00\%) & 1.00 (+0.00\%) & 1.00 (+0.00\%)  \\ \hline 
      $x$ $[-]$ & 1.00 & 1.00 (-0.00\%) & 1.00 (-0.00\%) & 1.00 (-0.00\%) & 1.00 (-0.00\%)  \\ \hline 
      $y$ $[-]$ & 1.00 & 1.00 (-0.00\%) & 1.00 (-0.00\%) & 1.00 (+0.00\%) & 1.00 (+0.00\%)  \\ \hline 
      Iterations   & - & 4.77 (0.00\%) & 4.68 (-1.87\%) & 6.91 (+44.76\%) & 7.04 (+47.52\%)  \\ \hline 
      Failures & - & 163 (16.30\%) & 250 (25.00\%) & 265 (26.50\%) & 437 (43.70\%)  \\ \hline 
  \end{tabular} 
 \end{center} 
\end{table} 
\FloatBarrier

Note that in this case a trust region was implemented on the first three iterations, bounding $|\Delta x_i|/x_i \leq [0.2,0.5,1.0]$.  In these early iterations, the Hessian approximation is quite poor and can send the candidate point off very far from the local region of interest.  In addition, all three algorithms reach 100\% success in the case of a good initial guess, but in Figure \ref{ros_b} between 15\% and 45\% of the trials converge to the local optimum $f(0,0) = 2$, which constitutes a failure of the algorithm.

So why does the traditional SQP outperform the LSQP modification?  Just because the problem can be constructed as a Signomial Program does not imply underlying log-convexity, and the Rosenbrock Problem exhibits almost no underlying log-convexity.  

\begin{figure}[htb]
\centering     
\subfigure[Rosenbrock Objective Function]{\includegraphics[width=0.48\textwidth]{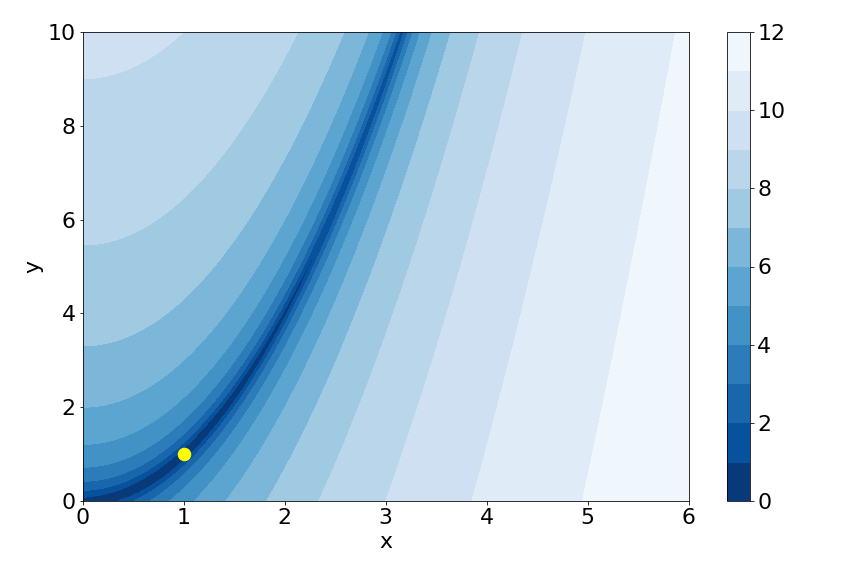} \label{ros_ut1}} 
\subfigure[Quadratic Approximation]{\includegraphics[width=0.48\textwidth]{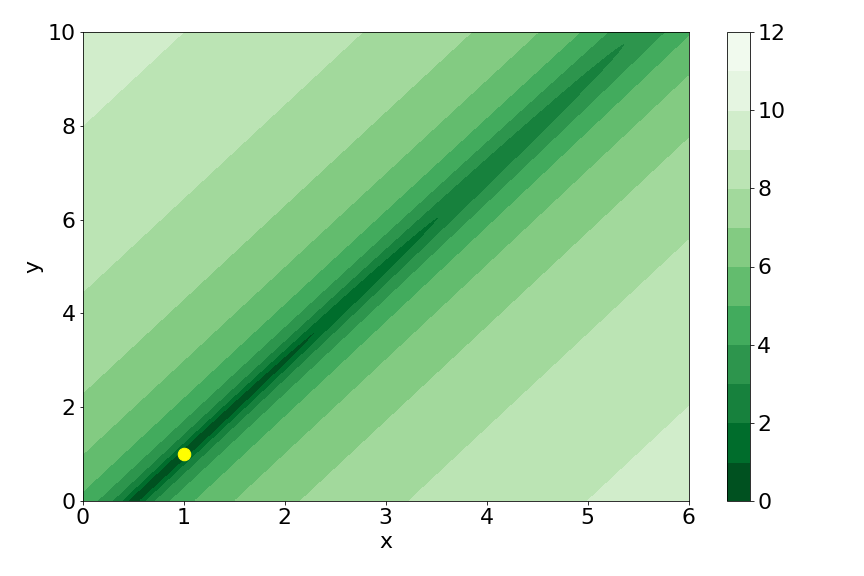} \label{ros_ut2}} 
\caption{A quadratic approximation of the Rosenbrock objective function in untransformed space.  Approximation referenced about the global optimum indicated by a yelllow dot.}
\label{ros_space1}
\end{figure}

\begin{figure}[htb]
\centering     
\subfigure[Log Transformation]{\includegraphics[width=0.48\textwidth]{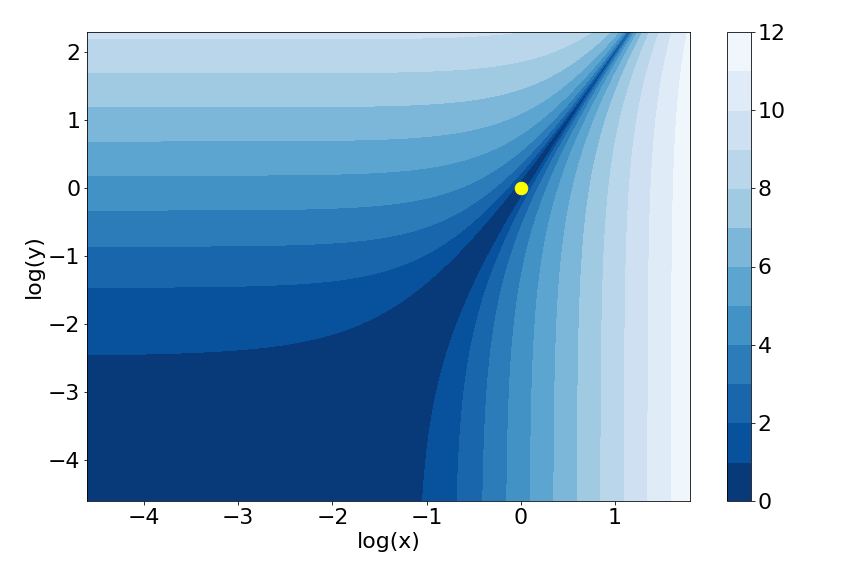} \label{ros_t1}} 
\subfigure[Quadratic Approximation]{\includegraphics[width=0.48\textwidth]{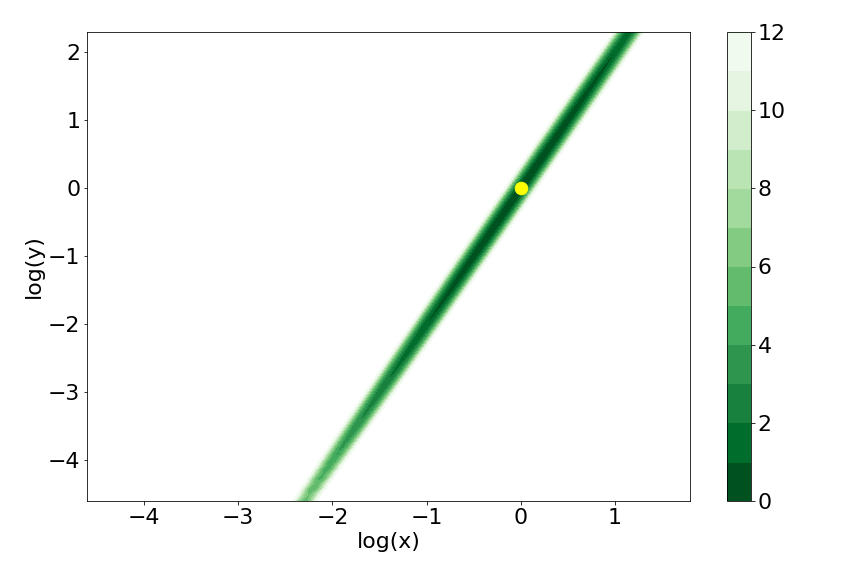} \label{ros_t2}} 
\caption{A quadratic approximation of the Rosenbrock objective function in log transformed space.  Approximation referenced about the global optimum indicated by a yellow dot.  Note the large white area in (b) represents a region where the value of the function exceeds the contour range.}
\label{ros_space2}
\end{figure}

This can be seen by visualizing the original Rosenbrock problem and the Rosenbrock problem under log transformation directly in Figures \ref{ros_space1} and \ref{ros_space2}.

\FloatBarrier

The traditional SQP algorithm takes QP approximations in the space of Figure \ref{ros_ut1}, resulting in a QP objective function that appears like Figure \ref{ros_ut2}.  Compare this visually to Figure \ref{ros_space2}, which represents the LSQP modification.  In order to capture the local region around the reference point, the quadratic approximation in Figure \ref{ros_t2} creates a deep narrow valley that clearly is a poor approximation of the objective function globally, indicated by the large white area where the contour range is exceeded.  In other words, there is generally far more agreement between the original objective function Figure \ref{ros_ut1} and its quadratic approximations Figure \ref{ros_ut2} than there is between the transformed objective Figure \ref{ros_t1} and its quadratic approximations Figure \ref{ros_t2}.  Since the traditional SQP algorithm has superior sub-problem representations, it is no surprise that it outperforms the LSQP modification in this case.

\subsection{Results from the Floudas Heat Exchanger Problem}
\begin{figure}[htb]
\centering     
\subfigure[A Good Initial Guess]{\includegraphics[width=0.48\textwidth]{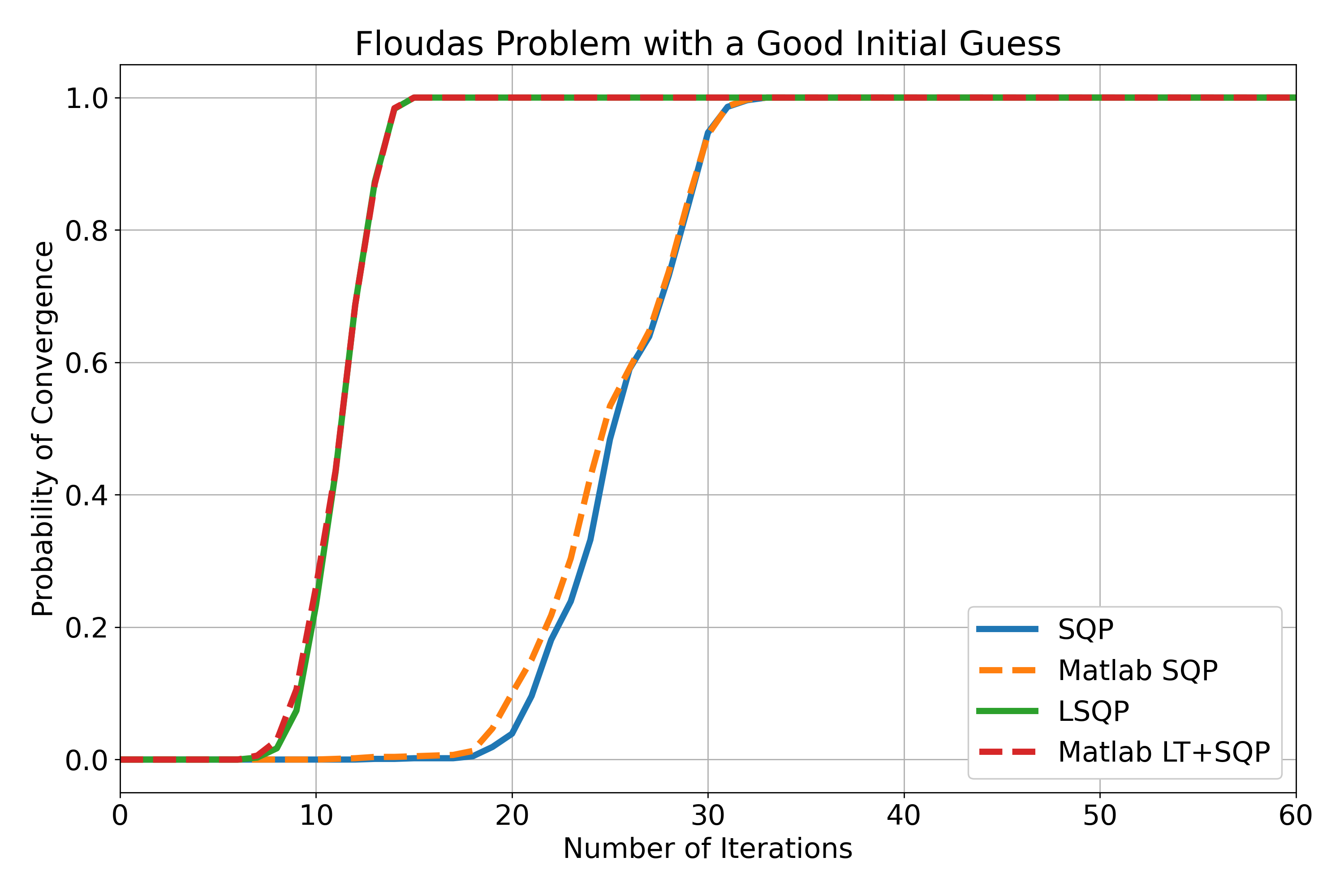} \label{fl_a}}
\subfigure[A Poor Initial Guess]{\includegraphics[width=0.48\textwidth]{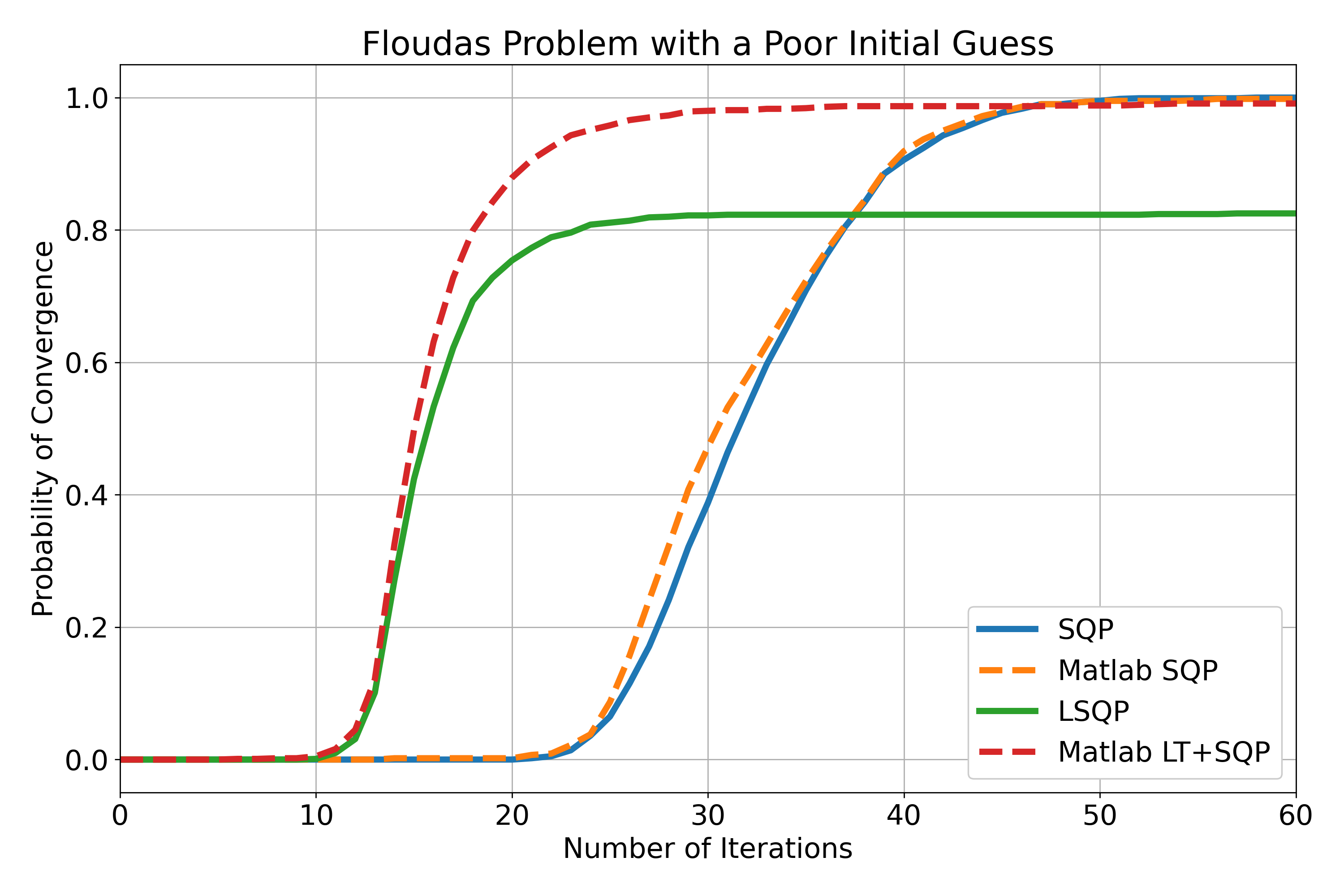} \label{fl_b}}
\caption{Probability of Convergence vs. Iteration Count for the Floudas Problem}
\label{floudas_results_fig}
\end{figure}

Floudas \cite{floudas2013handbook} provides the following example for the design of a heat exchanger:
\begin{equation}
    \begin{aligned}
         \underset{x_1,...,x_8}{\text{minimize}}  \quad &  x_1 + x_2 + x_3 \\
         \text{subject to}  \quad & \frac{833.33252 x_4}{x_2 x_6} + \frac{100}{x_6} - \frac{83333.333}{x_1 x_6} \leq 1 \\
         & \frac{1250 x_5}{x_2 x_7} + \frac{x_4}{x_7} - \frac{1250 x_4}{x_2 x_7} \leq 1 \\
         & \frac{1250000}{x_3 x_8} + \frac{x_5}{x_8} - \frac{2500 x_5}{x_3 x_8}  \leq 1 \\
         & 0.0025 x_4 + 0.0025 x_6  \leq 1 \\
         & -0.0025 x_4 + 0.0025 x_5 + 0.0025 x_7  \leq 1 \\
         & -0.01 x_5 + 0.01 x_8  \leq 1
    \end{aligned}
    \label{floudas}
\end{equation}
The problem has 5 signomial constraints, and only a single GP-compatible posynomial (the 4th constraint).  In contrast with the Rosenbrock Problem, which has signomials but little underlying log-convexity, the Floudas Problem does indeed exhibit some degree of log-convexity.  This underlying structure is apparent in the results presented in Figure \ref{fl_a} as the LSQP modification achieves 100\% success before reaching 20 iterations, while the SQP algorithms take twice as many iterations to achieve the same success rate.  
\FloatBarrier
\FloatBarrier
Figure \ref{fl_b} is a more nuanced result.  For the majority of trials where LSQP does converge, it converges in fewer iterations.  However 174 (17.4\%) of the LSQP trials failed.  These failures are a result of one or more of the constraints violating the conditions $g_i(\mathbf{x}_k) > 0$ and $h_j(\mathbf{x}_k) > 0$, thus causing the log transformation to fail at one of the iterations.  

A proposal to correct these failures in future work was mentioned in Section \ref{lsqp_algo}: the line search phase of Algorithm~\ref{alg:lsqp} (Line 12) can be bounded to keep all $g_i(\mathbf{x}_k) > 0$ and $h_j(\mathbf{x}_k) > 0$ at each step $k$.  Though not possible to confirm, it is likely that Matlab's SQP algorithm is doing this bounding in keeping all candidate points $x_k$ real, hence the higher success rate.  However, enforcing these bounds comes at the expense of extra calls to the objective and constraint functions, so the unmodified line search was implemented here specifically to demonstrate this failure mode.  Future work will correct this flaw.

\begin{table}[htbp] 
  \footnotesize 
  \begin{center} 
  \caption{Results of Solving the Floudas Problem With a Good Initial Guess} 
  \label{t:scheme_comparison_Floudas_Good} 
  \begin{tabular}{ c  c  c  c  c  c } 
      \cline{2-6}
                  & Optimum & SQP              & Matlab SQP        & LSQP              & Matlab LT+SQP       \\ 
      \hline 
      Obj $[-]$   & 7049.2 & 7049.2 (-0.00\%)  & 7049.2 (-0.00\%)  & 7049.2 (-0.00\%)  & 7049.2 (-0.00\%)  \\ \hline 
      $x_1$ $[-]$ & 579.3  & 579.3  (-0.00\%)  & 579.3  (-0.00\%)  & 579.3  (-0.00\%)  & 579.3  (-0.00\%)   \\ \hline 
      $x_2$ $[-]$ & 1360.0 & 1360.0 (-0.00\%)  & 1360.0 (-0.00\%)  & 1360.0 (-0.00\%)  & 1360.0 (+0.00\%)  \\ \hline 
      $x_3$ $[-]$ & 5110.0 & 5110.0 (-0.00\%)  & 5110.0 (-0.00\%)  & 5110.0 (-0.00\%)  & 5110.0 (-0.00\%)  \\ \hline 
      $x_4$ $[-]$ & 182.0  & 182.0  (-0.00\%)  & 182.0  (-0.00\%)  & 182.0  (-0.00\%)  & 182.0  (-0.00\%)   \\ \hline 
      $x_5$ $[-]$ & 295.6  & 295.6  (+0.00\%)  & 295.6  (+0.00\%)  & 295.6  (+0.00\%)  & 295.6  (+0.00\%)   \\ \hline 
      $x_6$ $[-]$ & 218.0  & 218.0  (+0.00\%)  & 218.0  (+0.00\%)  & 218.0  (+0.00\%)  & 218.0  (+0.00\%)   \\ \hline 
      $x_7$ $[-]$ & 286.4  & 286.4  (-0.00\%)  & 286.4  (-0.00\%)  & 286.4  (-0.00\%)  & 286.4  (-0.00\%)   \\ \hline 
      $x_8$ $[-]$ & 395.6  & 395.6  (+0.00\%)  & 395.6  (+0.00\%)  & 395.6  (+0.00\%)  & 395.6  (+0.00\%)   \\ \hline 
      Iterations  & -      & 25.87  (0.00\%)   & 25.42  (-1.71\%)  & 11.70 (-54.79\%)  & 11.62 (-55.08\%)    \\ \hline 
      Failures    & -      & 0 (0.00\%)        & 0 (0.00\%)        & 0 (0.00\%)        & 0 (0.00\%)          \\ \hline 
  \end{tabular} 
 \end{center} 
\end{table} 

\begin{table}[htbp] 
  \footnotesize 
  \begin{center} 
  \caption{Results of Solving the Floudas Problem With a Poor Initial Guess} 
  \label{t:scheme_comparison_Floudas_Poor} 
  \begin{tabular}{ c  c  c  c  c  c } 
      \cline{2-6}
                  & Optimum  & SQP                & Matlab SQP         & LSQP               & Matlab LT+SQP       \\ 
      \hline 
      Obj $[-]$   & 7049.2 & 7049.2  (-0.00\%) & 7049.2 (-0.00\%) & 7050.8 (+0.02\%) & 7049.4 (+0.00\%)  \\ \hline 
      $x_1$ $[-]$ & 579.3  & 579.3  (-0.00\%)  & 579.3 (-0.00\%)  & 577.9 (-0.24\%)  & 578.8 (-0.09\%)   \\ \hline 
      $x_2$ $[-]$ & 1360.0 & 1360.0  (+0.00\%) & 1360.0 (+0.00\%) & 1358.7 (-0.10\%) & 1360.4 (+0.03\%)  \\ \hline 
      $x_3$ $[-]$ & 5110.0 & 5110.0  (-0.00\%) & 5110.0 (-0.00\%) & 5114.3 (+0.08\%) & 5110.2 (+0.00\%)  \\ \hline 
      $x_4$ $[-]$ & 182.0  & 182.0  (-0.00\%)  & 182.0 (-0.00\%)  & 181.9 (-0.09\%)  & 181.9 (-0.08\%)   \\ \hline 
      $x_5$ $[-]$ & 295.6  & 295.6  (+0.00\%)  & 295.6 (+0.00\%)  & 295.5 (-0.05\%)  & 295.3 (-0.09\%)   \\ \hline 
      $x_6$ $[-]$ & 218.0  & 218.0  (+0.00\%)  & 218.0 (+0.00\%)  & 218.1 (+0.07\%)  & 217.9 (-0.05\%)   \\ \hline 
      $x_7$ $[-]$ & 286.4  & 286.4  (-0.00\%)  & 286.4 (-0.00\%)  & 286.4 (-0.00\%)  & 286.1 (-0.10\%)   \\ \hline 
      $x_8$ $[-]$ & 395.6  & 395.6  (+0.00\%)  & 395.6 (+0.00\%)  & 395.5 (-0.04\%)  & 395.7 (+0.03\%)   \\ \hline 
      Iterations  & -        & 32.79 (0.00\%)     & 32.22 (-1.73\%)    & 16.29 (-50.32\%)   & 16.64 (-49.26\%)    \\ \hline 
      Failures    & -        & 0 (0.00\%)         & 0 (0.00\%)         & 174 (17.40\%)      & 8 (0.80\%)          \\ \hline 
  \end{tabular} 
 \end{center} 
\end{table} 

On the whole, the Floudas problem is another win for LSQP, as significant computational savings is achieved by implementing the log transformation.  Work remains to be done in the robustness of the LSQP algorithm as presented here (see Section \ref{lsqp_algo}), but the successful exploitation of the underlying problem structure is clear.
\subsection{Results from the Kirschen-Ozturk Aircraft Design Problem} \hfill \break
Kirschen \cite{kirschen2018power} proposes a signomial program representing a low fidelity aircraft sizing, originally attributed to Ozturk.  This problem is an extension of one of the more simple cases proposed by Hoburg \cite{hoburg2014geometric}.  It is GP compatible with the exception of a single fuel volume constraint, making it a non-convex Signomial Program.  It has been reformulated here to conform to LSQP standard form:
\begin{align*}
    \text{minimize}  \quad &  W_f \\
    \text{subject to}  \quad & \frac{c_T t D}{W_f} \leq 1  \\
    & \frac{R}{V t} \leq 1 \\
    & \frac{\frac12 \rho V^2 S C_D}{D} \leq 1 \\
    & \frac{A_{C_{D_0}}}{S C_D} + \frac{k C_f}{C_D} \frac{S_{wet}}{S} + \frac{C_L^2}{\pi A e C_D} \leq 1 \\
    & \frac{0.074 Re^{-0.02}}{C_f} \leq 1 \\
    & \frac{\mu Re}{\rho V \sqrt{S/A}} \leq 1  \\
    & \frac{W_0}{\frac12 \rho V^2 S C_L} + \frac{W_w}{\frac12 \rho V^2 S C_L} + \frac{\frac12 W_f }{\frac12 \rho V^2 S C_L}  \leq 1 \\
    & \frac{W}{\frac12 \rho V_{min}^2 S C_{L_{max}}} \leq 1 \tag{\stepcounter{equation}\theequation}\\
    & \frac{W_0}{W} + \frac{W_w}{W} + \frac{W_f}{W} \leq 1 \\
    & \frac{W_{w_{surf}}}{W_w} + \frac{W_{w_{strc}}}{W_w} \leq 1 \\
    & \frac{C_{W_{w,1}} S}{W_{w_{surf}}} \leq 1 \\
    & C_{W_{w,2}} \frac{N_{ult} A^{\frac32} \sqrt{(W_0 + V_{f_{fuse}} g \rho_f) W S}}{W_{w_{strc}} \tau} \leq 1 \\
    & \frac{V_f}{V_{f_{avail}}} \leq 1 \\
    & \frac{V_f g \rho_f}{W_f} = 1 \\
    & V_{f_{avail}} - V_{f_{wing}} - V_{f_{fuse}} + 1 \leq 1  \\
    & \frac{V_{f_{wing}}^2 A}{0.0009 S^3 \tau^2} \leq 1 \\
    & \frac{V_{f_{fuse}}}{A_{C_{D_0}} 10[m] } \leq  1 
\end{align*}
Variables were also constrained to be greater than a small positive constant in order to assist in the construction of sub-problems.  Note that the signomial constraint $V_{f_{avail}} - V_{f_{wing}} - V_{f_{fuse}} + 1 \leq 1$ was created using a non-ideal construction method in order to provide the greatest challenge to LSQP.

Unlike the previous problems which had known optima, this problem is non-convex and has no known solution \textit{a priori}.  Thus, the reference solution in this case is determined by using a Signomial Programming formulation and the Difference of Convex Algorithm (abbreviated as SP+DCA).  This solution method is used in the original publication \cite{kirschen2018power} and will therefore be treated as the optimum for the purposes of this section.

\begin{figure}[htb]
\centering     
\subfigure[A Good Initial Guess]{\includegraphics[width=0.48\textwidth]{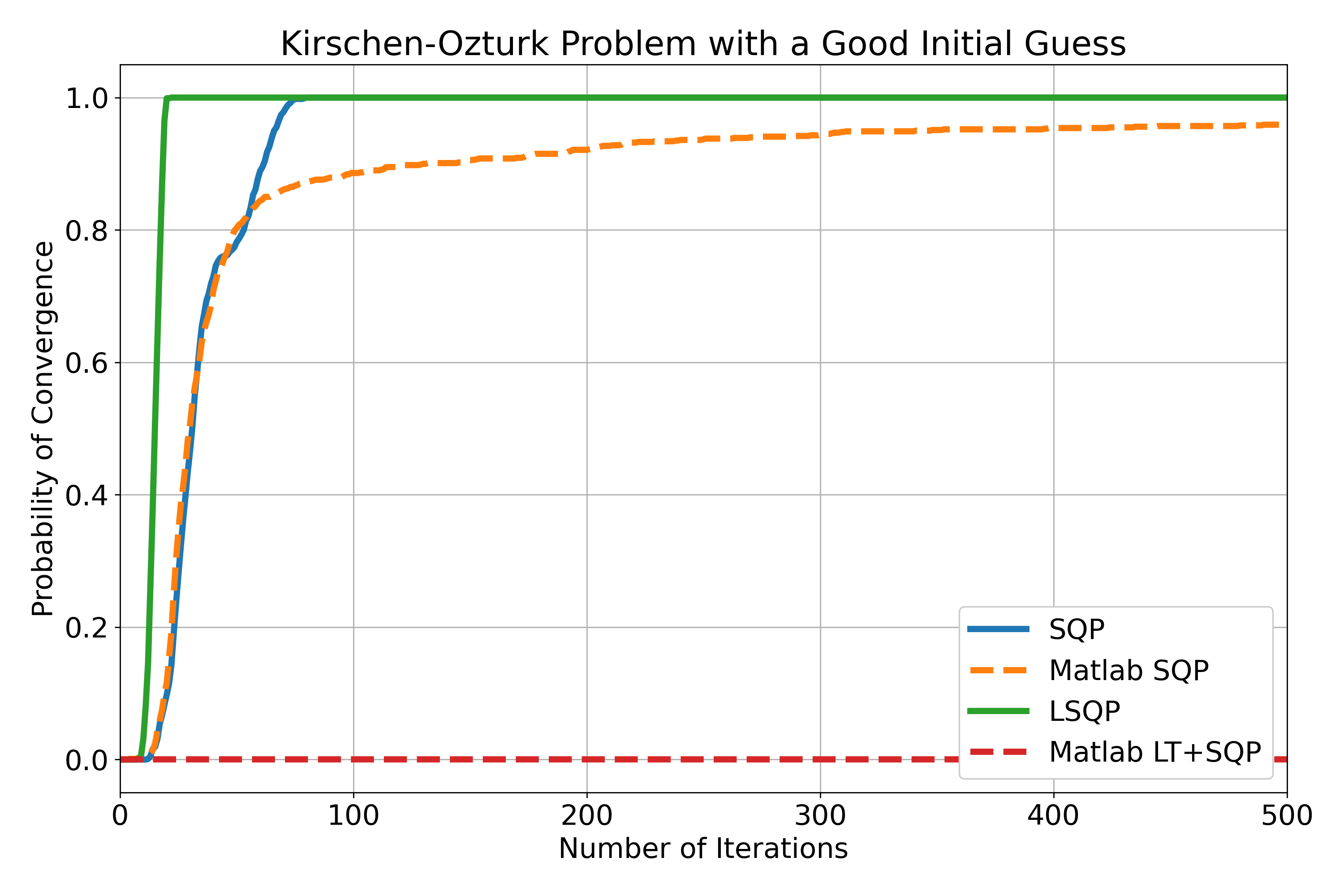} \label{ko_a}}
\subfigure[A Poor Initial Guess]{\includegraphics[width=0.48\textwidth]{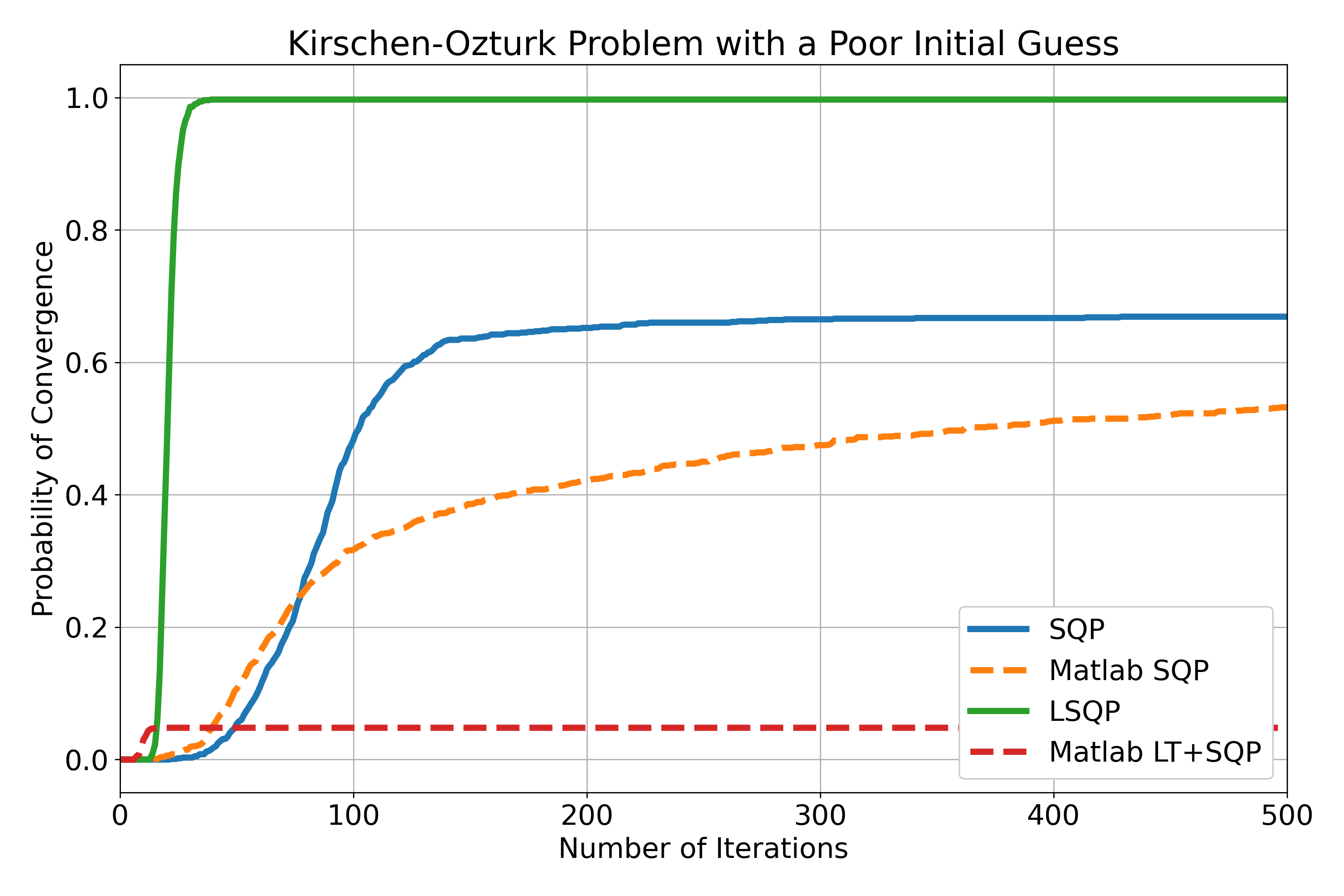} \label{ko_b}}
\caption{Probability of Convergence vs. Iteration Count for the Kirschen-Ozturk Problem}
\label{kop}
\end{figure}

Kirschen showed an approximately 40\% decrease in the number of iterations between the log transformed problem and the original problem \cite{kirschen2018power}, but only considered a single initial guess that yielded an interpretable result.  Figure \ref{ko_a} and Table \ref{t:scheme_comparison_Kirschen-Ozturk_Good} confirm that result with a good initial guess.  However the trials with poor initial guesses demonstrate a far more significant win for LSQP, as shown in Table \ref{t:scheme_comparison_Kirschen-Ozturk_Poor} and Figure \ref{ko_b}. 

Figure \ref{kop} also shows that almost none of the Matlab LT+SQP cases converged.  Further analysis revealed that in these cases the algorithm terminated at an infeasible point due to a shrinking step size, but no further information is provided by the Matlab output.  Those cases that did converge were quite far from the known optimum and are therefore not reported.  It is difficult to draw a general conclusion from this single test problem, but the failure of the Matlab LT+SQP on so many cases further points to the value of a dedicated LSQP algorithm as opposed to working with an existing SQP solver.

\begin{table}[htbp] 
  \footnotesize 
  \begin{center} 
  \caption{Results of Solving the Kirschen-Ozturk Problem With a Good Initial Guess} 
  \label{t:scheme_comparison_Kirschen-Ozturk_Good} 
  \begin{tabular}{ c  c  c  c  c  c } 
      \cline{2-5}
                              & SP+DCA          & SQP                 & Matlab SQP          & LSQP                \\ 
      \hline 
      Obj $[N]$               & 755.91          & 756.28 (+0.05\%)    & 756.64 (+0.10\%)    & 755.90 (-0.00\%)   \\ \hline 
      $A$ $[-]$               & 6.52            & 6.46 (-0.89\%)      & 6.55 (+0.50\%)      & 6.51 (-0.11\%)     \\ \hline 
      $C_{D}$ $[-]$           & 0.013           & 0.013 (-0.17\%)     & 0.013 (+0.14\%)     & 0.013 (-0.01\%)     \\ \hline 
      $C_{L}$ $[-]$           & 0.234           & 0.233 (-0.65\%)     & 0.235 (+0.53\%)     & 0.234 (-0.06\%)     \\ \hline 
      $C_{f}$ $[-]$           & 3.278e-03       & 3.273e-03 (-0.16\%) & 3.280e-03 (+0.05\%) & 3.277e-03 (-0.01\%) \\ \hline 
      $D$ $[N]$               & 368.7           & 370.3 (+0.45\%)     & 368.5 (-0.03\%)     & 368.8 (+0.03\%)   \\ \hline 
      $Re$ $[-]$              & 5.86e+06        & 5.92e+06 (+0.91\%)  & 5.86e+06 (-0.06\%)  & 5.869e+06 (+0.07\%) \\ \hline 
      $S$ $[m^2]$             & 16.00           & 15.97 (-0.14\%)     & 16.02 (+0.15\%)     & 15.99 (-0.02\%)    \\ \hline 
      $V$ $[m/s]$             & 54.18           & 54.40 (+0.40\%)     & 54.12 (-0.13\%)     & 54.21 (+0.03\%)    \\ \hline 
      $V_{f}$ $[m^3]$         & 0.096           & 0.096 (+0.05\%)     & 0.096 (+0.10\%)     & 0.096 (-0.00\%)     \\ \hline 
      $V_{f_{avail}}$ $[m^3]$ & 9.584e-02       & 9.589e-02 (+0.05\%) & 9.593e-02 (+0.10\%) & 9.584e-02 (-0.00\%) \\ \hline 
      $V_{f_{fuse}}$ $[m^3]$  & 5.840e-03       & 5.454e-03 (-6.61\%) & 5.858e-03 (+0.30\%) & 5.656e-03 (-3.15\%) \\ \hline 
      $V_{f_{wing}}$ $[m^3]$  & 9.016e-02       & 9.043e-02 (+0.30\%) & 9.016e-02 (+0.00\%) & 9.018e-02 (+0.02\%) \\ \hline 
      $W$ $[N]$               & 7140.2          & 7130.1 (-0.14\%)    & 7151.0 (+0.15\%)    & 7138.6 (-0.02\%)  \\ \hline 
      $W_{f}$ $[N]$           & 755.9           & 756.3 (+0.05\%)     & 756.6 (+0.10\%)     & 755.9 (-0.00\%)   \\ \hline 
      $W_{w}$ $[N]$           & 1444.3          & 1433.8 (-0.73\%)    & 1454.4 (+0.70\%)    & 1442.7 (-0.11\%)  \\ \hline 
      $W_{w_{strc}}$ $[N]$    & 720.8           & 711.3 (-1.32\%)     & 729.8 (+1.24\%)     & 719.4 (-0.20\%)   \\ \hline 
      $W_{w_{surf}}$ $[N]$    & 723.5           & 722.4 (-0.14\%)     & 724.6 (+0.15\%)     & 723.3 (-0.02\%)   \\ \hline 
      $t$ $[min]$             & 307.6           & 306.5 (-0.35\%)     & 308.2 (+0.22\%)     & 307.5 (-0.03\%)   \\ \hline 
      Iterations              & -               & 35.93 (0.00\%)      & 47.67 (+32.68\%)    & 15.36 (-57.24\%)    \\ \hline 
      Failures                & -               & 0 (0.00\%)          & 39 (3.90\%)         & 0 (0.00\%)          \\ \hline 
  \end{tabular} 
 \end{center} 
\end{table}

\begin{table}[htbp] 
  \footnotesize 
  \begin{center} 
  \caption{Results of Solving the Kirschen-Ozturk Problem With a Poor Initial Guess} 
  \label{t:scheme_comparison_Kirschen-Ozturk_Poor} 
  \begin{tabular}{ c  c  c  c  c  c } 
      \cline{2-5}
                              & SP+DCA    & SQP                   & Matlab SQP            & LSQP                 \\ 
      \hline 
      Obj $[N]$               & 755.91    & 761.50 (+0.74\%)      & 782.30 (+3.49\%)      & 755.90 (-0.00\%)    \\ \hline 
      $A$ $[-]$               & 6.52      & 6.02 (-7.71\%)        & 6.06 (-7.11\%)        & 6.51 (-0.11\%)      \\ \hline 
      $C_{D}$ $[-]$           & 0.013     & 0.013 (-1.59\%)       & 0.013 (-1.28\%)       & 0.013 (-0.01\%)      \\ \hline 
      $C_{L}$ $[-]$           & 0.234     & 0.220 (-5.95\%)       & 0.223 (-4.75\%)       & 0.234 (-0.06\%)      \\ \hline 
      $C_{f}$ $[-]$           & 3.278e-03 & 3.229e-03 (-1.50\%)   & 3.305e-03 (+0.83\%)   & 3.277e-03 (-0.01\%)  \\ \hline 
      $D$ $[N]$               & 368.7     & 386.5 (+4.48\%)      & 458.6 (+24.40\%)      & 368.8 (+0.03\%)    \\ \hline 
      $Re$ $[-]$              & 5.86e+06  & 6.447e+06 (+9.93\%)   & 6.254e+06 (+6.65\%)   & 5.869e+06 (+0.07\%)  \\ \hline 
      $S$ $[m^2]$             & 16.00     & 15.81 (-1.12\%)     & 64.82 (+305.33\%)     & 15.99 (-0.02\%)     \\ \hline 
      $V$ $[m/s]$             & 54.18     & 56.34 (+3.96\%)     & 117.10 (+116.10\%)    & 54.21 (+0.03\%)     \\ \hline 
      $V_{f}$ $[m^3]$         & 0.096     & 0.097 (+0.74\%)     & 0.486 (+406.63\%)     & 0.096 (-0.00\%)      \\ \hline 
      $V_{f_{avail}}$ $[m^3]$ & 9.584e-02 & 9.680e-02 (+1.00\%) & 4.919e-01 (+413.28\%) & 9.584e-02 (-0.00\%)  \\ \hline 
      $V_{f_{fuse}}$ $[m^3]$  & 5.840e-03 & 4.535e-03 (-22.35\%)  & 1.023e-02 (+75.11\%)  & 5.656e-03 (-3.15\%)  \\ \hline 
      $V_{f_{wing}}$ $[m^3]$  & 9.016e-02 & 9.272e-02 (+2.83\%) & 8.456e-01 (+837.91\%) & 9.018e-02 (+0.02\%)  \\ \hline 
      $W$ $[N]$               & 7140.2    & 7060.0 (-1.12\%)     & 9454.5 (+32.41\%)     & 7138.6 (-0.02\%)   \\ \hline 
      $W_{f}$ $[N]$           & 755.9     & 761.5 (+0.74\%)       & 782.3 (+3.49\%)       & 755.9 (-0.00\%)    \\ \hline 
      $W_{w}$ $[N]$           & 1444.3    & 1358.5 (-5.94\%)    & 3702.4 (+156.35\%)    & 1442.7 (-0.11\%)   \\ \hline 
      $W_{w_{strc}}$ $[N]$    & 720.8     & 643.2 (-10.77\%)       & 763.3 (+5.89\%)       & 719.4 (-0.20\%)    \\ \hline 
      $W_{w_{surf}}$ $[N]$    & 723.5     & 715.3 (-1.12\%)     & 2938.0 (+306.09\%)    & 723.3 (-0.02\%)    \\ \hline 
      $t$ $[min]$             & 307.6     & 297.6 (-3.24\%)       & 297.7 (-3.22\%)       & 307.5 (-0.03\%)    \\ \hline 
      Iterations              & -         & 91.18 (0.00\%)        & 129.82 (+42.38\%)     & 21.06 (-76.90\%)     \\ \hline 
      Failures                & -         & 331 (33.10\%)         & 468 (46.80\%)         & 3 (0.30\%)           \\ \hline 
  \end{tabular} 
 \end{center} 
\end{table}

This test problem is another clear win for the LSQP modification, and is perhaps the most significant result of this work.  The best result is shown in Figure \ref{ko_b}: an aircraft design problem is solved with a 74\% reduction in number of iterations and with a 45\% higher success rate when using the LSQP modification as opposed to traditional SQP.  
\subsection{When Should LSQP be Used Over SQP?}
In light of the evidence presented here, the question remains when LSQP should be used in place of SQP.  First, three tenants must hold true:  variables are expected to be strictly positive, the objective function is expected to be strictly positive, and constraints in LSQP standard form are expected to be strictly positive.  If these tenants hold, then LSQP should be considered.  In the case of the Floudas problem, these tenants alone are sufficient to show a significant improvement in the number of iterations required for convergence.  But as was shown in the Rosenbrock problem, these tenants can hold true and still not imply log-convexity.  Thus, one additional indicator for use of LSQP is a significant percentage of GP-compatible constraints (monomials and posynomials).  If GP compatible constraints are present in large numbers, then some degree of log-convexity can be expected, and LSQP should be strongly considered.
\section{Conclusions}
This work demonstrates that LSQP solves some engineering design problems faster than classical SQP, and is capable of solving problems that are not solvable by classical SQP.  Thus, LSQP is a new tool that both improves and extends existing capability.

While previous research efforts in Geometric and Signomial Programming have exposed the log-convex structure present in many engineering design problems, the adoption of these methodologies has been slow.  To adopt a GP or SP approach requires total commitment to a new process and a rewriting of existing models, a barrier too substantial for most designers.  LSQP provides a middle ground, enabling the exploitation of log-convexity while requiring no change to existing processes or discarding of trusted black box models.  As a result of the work done here, LSQP can be viewed as a direct substitute for traditional SQP and the potential computational savings obtained from this simple switch are significant. 

\section*{Funding Sources}
This material is based on research sponsored by the U.S. Air Force under agreement number FA8650-20-2-2002. The U.S. Government is authorized to reproduce and distribute reprints for Governmental purposes notwithstanding any copyright notation thereon.  The views and conclusions contained herein are those of the authors and should not be interpreted as necessarily representing the official policies or endorsements, either expressed or implied, of the U.S. Air Force or the U.S. Government.

\section*{Acknowledgments}
The author would like to thank Bob Haimes and Mark Drela for their mentorship and support and for their notes on this paper, the EnCAPS Technical Monitor Ryan Durscher, Philippe Kirschen who laid some of the foundation for this work, Berk Ozturk who originally authored the fourth test problem, Devon Jedamski for his review and comments, and the two anonymous reviewers who provided comments and suggestions. 

The author also acknowledges the MIT SuperCloud and Lincoln Laboratory Supercomputing Center for providing HPC, database, and consultation resources that have contributed to the research results reported within this paper.

\bibliography{short_deck}
\end{document}